\documentclass{elsart}
\usepackage{graphicx}
\usepackage[dvips]{color}
\begin{document}

\begin{frontmatter}
\title{A comparison between several correlated stochastic volatility models}
\author[ub]{Josep Perell\'o\corauthref{cor1}}, 
\author[ub]{Jaume Masoliver},
\author[ub,gaesco]{and Napole\'on Anento}
\address[ub]{Departament de F\'{\i}sica Fonamental, Universitat de Barcelona, Diagonal, 647, E-08028 Barcelona, Spain}
\address[gaesco]{Gaesco Bolsa, SVB, S.A., Diagonal, 429, 08036-Barcelona,
Spain}
\corauth[cor1]{Corresponding author: josep.perello@ub.edu}
\date{\today}

\begin{abstract}
We compare the most common SV models such as the Ornstein-Uhlenbeck (OU), the Heston and the exponential OU (expOU) models. We try to decide which is the most appropriate one by studying their volatility autocorrelation and leverage effect, and thus outline the limitations of each model. We  add empirical research on market indices confirming the universality of the leverage and volatility correlations.
\end{abstract}
\maketitle
\end{frontmatter}

The multiplicative diffusion process is considered to be the most popular model in finance. We thus have that the log-price change without average describes the diffusive process $\dot{X}(t)=\sigma \xi_1(t)$ (in the It\^o sense), where $\sigma$ is the volatility (assumed to be constant) and $\xi_1$ is Brownian noise with $\langle\xi_1(t)\xi_1(t')\rangle=\delta(t-t')$. However, this model is unable to capture most of the statistical properties of real markets, those called stylized facts.

The stochastic volatility (SV) models are a natural way out to avoid the inconsistencies of the log-Brownian model by still assuming that $X$ follows a diffusion process but now with a random $\sigma$. However, the dynamics of $\sigma$ has not been definitively attached to any specific SV model. In this sense, we want to discern between the most common SV models. Among others we have:

(a) The Ornstein-Uhlenbeck (OU) SV model~\cite{stein91,masoli}
\begin{equation}
\dot{\sigma}(t)=-\alpha(\sigma-m)+k\xi_2(t), \qquad \sigma(t)=m+k\int_{-\infty}^t e^{-\alpha(t-t')}\xi_2(t')dt'.
\label{dou}
\end{equation}
(b) The Heston SV model~\cite{heston,dragulescu} (with $V\equiv\sigma^2$)
\begin{eqnarray}
&&\dot{V}(t)=-\alpha (V-m^2)+k\sqrt{V}\xi_2(t),\\
&&V(t)=m^2+k\int_{-\infty}^{t} e^{-\alpha(t-t')}\sqrt{V(t')}\xi_2(t')dt'.
\label{dheston}
\end{eqnarray}
(c) And, finally, the exponential Ornstein-Uhlenbeck (expOU) SV model~\cite{fouque2000} (with $Y\equiv\ln(\sigma/m)$)
\begin{equation}
\dot{Y}(t)=-\alpha Y+\frac{k}{m}\xi_2(t), \quad Y(t)=m+\frac{k}{m}\int_{-\infty}^t e^{-\alpha(t-t')}\xi_2(t')dt'.
\label{dexpou}
\end{equation}
Note that the Heston model does not provide a closed equation for the volatility. This can be a serious drawback in case we want to deal with analytical expressions. In addition, we suppose that noises are correlated (i.e., $\langle\xi_1(t) \xi_2(t')\rangle= \rho\delta(t-t')$ with $-1\leq\rho\leq 1$). This allows us to explain some stylized facts.

\begin{figure}[t]
\begin{center}
\includegraphics[angle=-90,width=6.25cm]{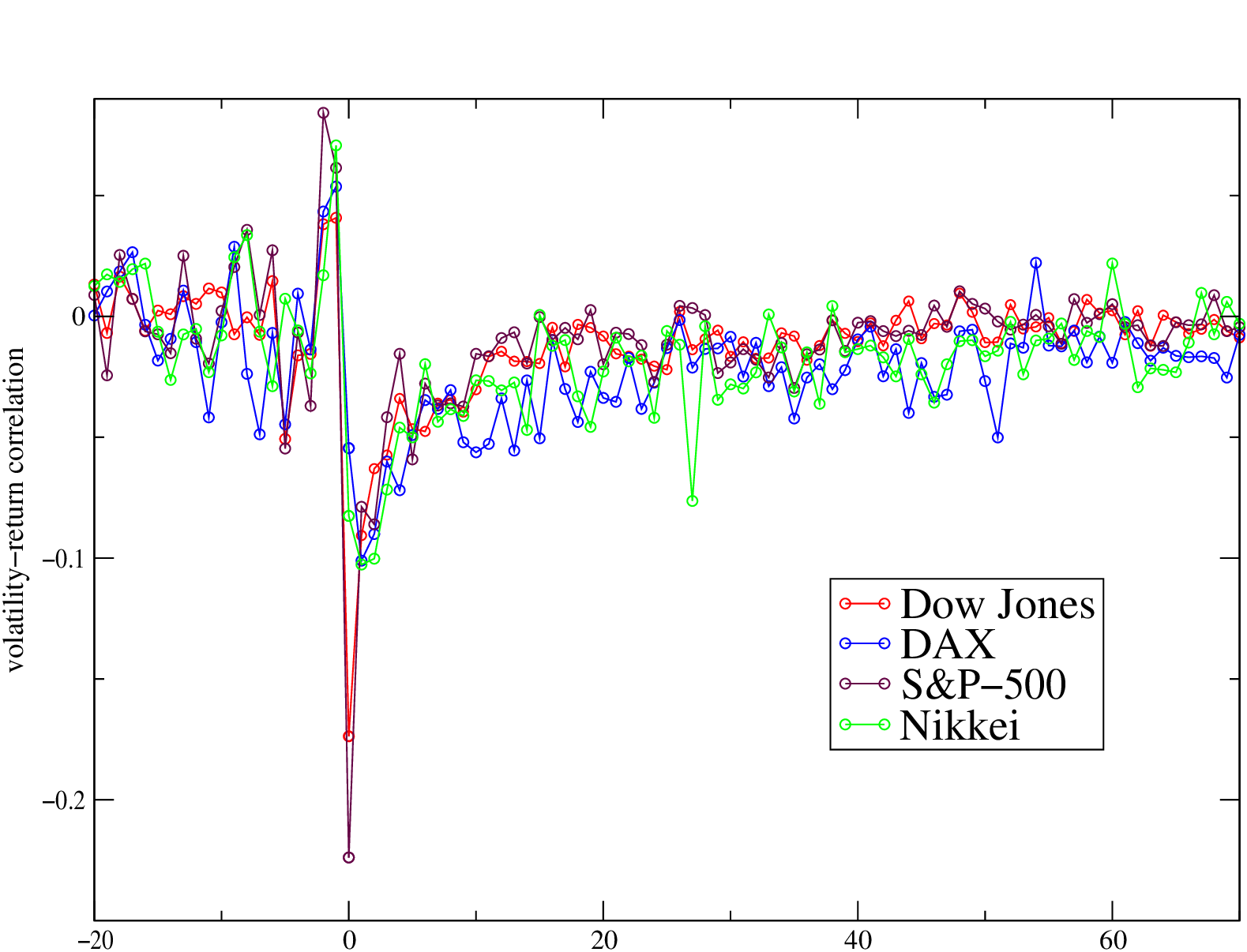} \hspace{0.1cm}
\includegraphics[angle=-90,width=6.5cm]{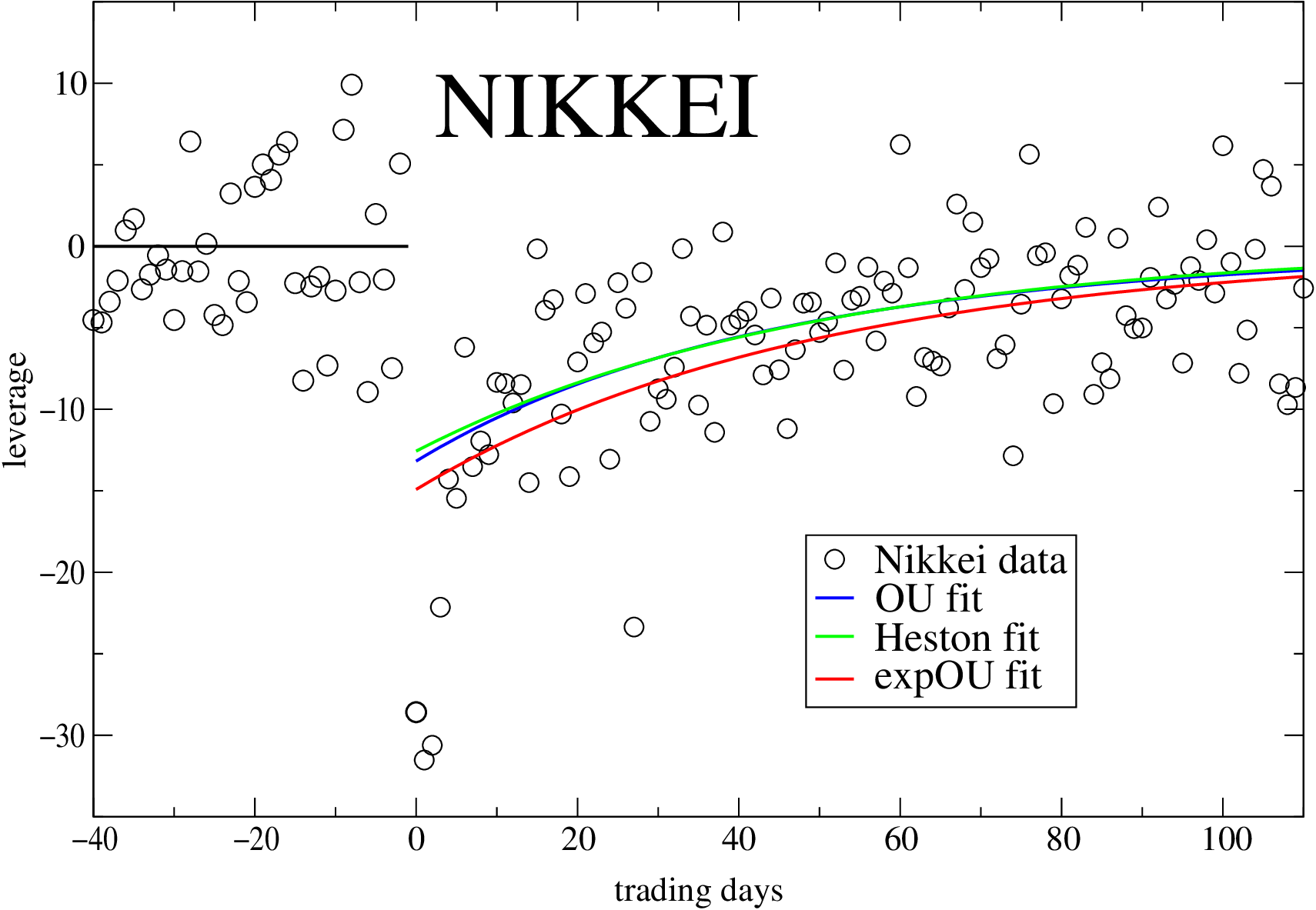}\\
\includegraphics[angle=-90,width=6.5cm]{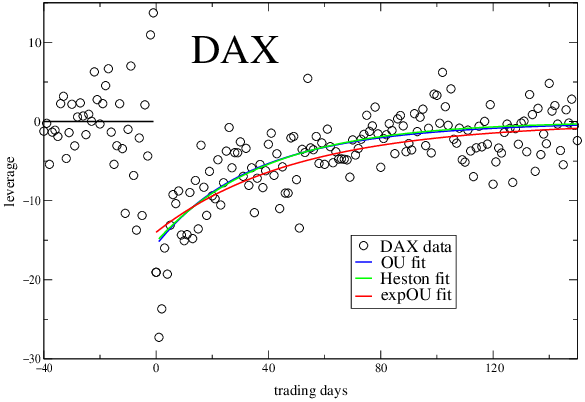}
\includegraphics[angle=-90,width=6.5cm]{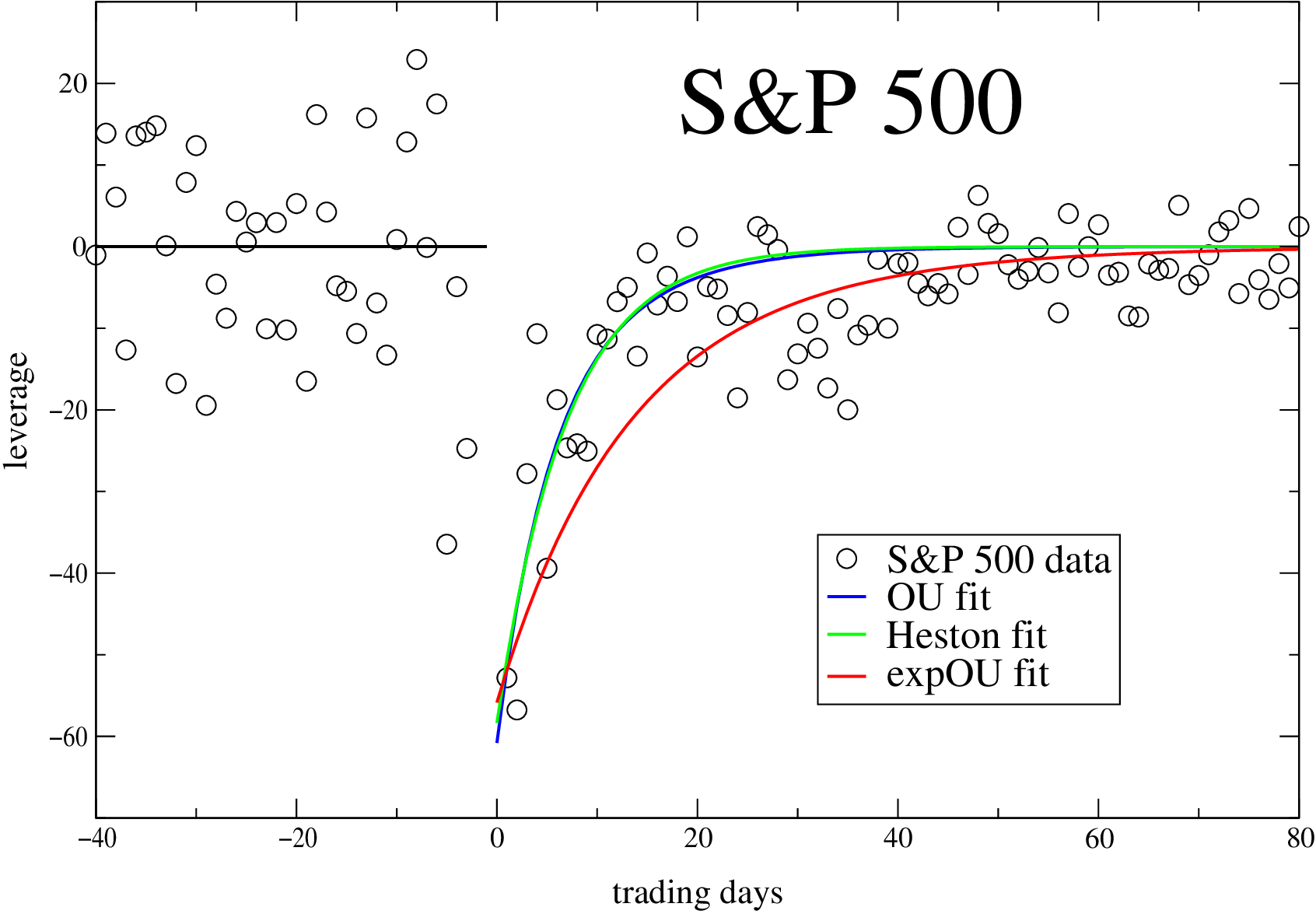}
\end{center}
\caption{\small The leverage effect for several daily price indices. We also add the leverage function $L(\tau)$ fit for the different SV models. The S\&P 500 is much more liquid.}
\label{leverage}
\end{figure}

The leverage effect~\cite{bouchPRL} is measured by:
$L(\tau)\equiv \langle dX(t+\tau)^2dX(t)\rangle/\langle dX(t)^2\rangle$. For the models above we have
\begin{eqnarray}
&&L_{\rm OU}(\tau)=2\rho\left[\frac{\nu\sqrt{2\alpha}\left(1+\nu^2e^{-\alpha\tau}\right)}{(1+\nu^2)^2m}\right] e^{-\alpha \tau}, \qquad
L_{\rm H}(\tau)\sim 4\rho\frac{\alpha}{k}e^{-\alpha\tau},
\label{lev}
\\
&&L_{\rm expOU}(\tau)=\frac{2\rho k}{m^2}e^{\nu^2/2}\exp{\left(2\nu^2e^{-\alpha\tau}-\alpha\tau\right)},
\label{lev2}
\end{eqnarray}
for $\tau>0$ and $L(\tau)=0$ for negative $\tau$. It seems impossible to derive an analytical expression for the Heston leverage because $V(t)$ does not have a closed expression (cf. Eq.~(\ref{dheston})). The leverage given in Eq.~(\ref{lev}) is exact only when $m^2=k^2/2\alpha$ but Ref.~\cite{dragulescu} shows that this is not true for the Dow Jones. We will have thus to simulate the process to compute the leverage effect in each specific Heston case. Eqs.~(\ref{lev})--(\ref{lev2}) are plotted and compared  with data in Fig.~\ref{leverage}. The data is too noisy to assert which is the most appropriate model. In any case, the three models have the desired exponential decay with a characteristic time of the order of few days. We also note that, without the correlation coefficient $\rho$, there is no leverage and data confirms that $\rho<0$~\cite{pre}.

\begin{figure}[tbp]
\begin{center}
\includegraphics[angle=-90, width=12cm]{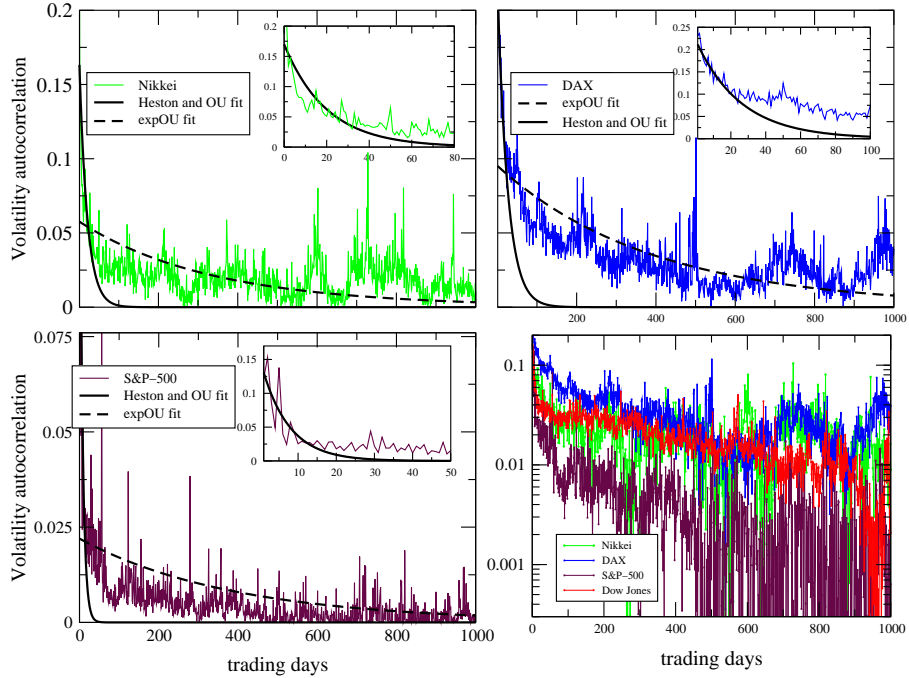}
\end{center}
\caption{The volatility correlation for several daily indices. We add two exponential fits one with a  short range (similar to the leverage time scale) and a second one for the larger time lags.}
\label{volauto}
\end{figure}

The volatility autocorrelation is  
$C(\tau)\equiv[\langle dX(t)^2dX(t+\tau)^2\rangle-\langle dX(t)^2\rangle\langle dX(t+\tau)^2\rangle]/\mbox{Var}[dX(t)^2]$. 
In this case, we can exactly derive this correlation for each model. We have
\begin{eqnarray}
&&C_{\rm OU}(\tau)=\frac{\nu^2e^{-\alpha\tau}(\nu^2e^{-\alpha\tau}+1)}{4\nu^2(2+\nu^2)+1},
\qquad C_{\rm H}(\tau)= \nu^2 e^{-\alpha\tau},\\
&&C_{\rm expOU}(\tau)= \frac{\exp{\left(4\nu^2e^{-\alpha\tau}\right)}-1}{3\exp(4\nu^2)-1}.
\end{eqnarray}
The volatility autocorrelation observed in empirical data shows at least two different time scales (see Fig.~\ref{volauto} and Ref.~\cite{pm}). The shortest one coincides with the leverage correlation time while the second scale is of the order of years, around 10 times bigger than the leverage time. The OU and Heston models are not capable of reproducing the slowest time decay but the expOU model has a nontrivial decay with the two time scales: 
\begin{equation}
C_{\rm expOU}(\tau)\sim e^{-\alpha\tau} \quad(\alpha\tau> 1), \qquad C_{\rm expOU}(\tau)\sim e^{-(k/m)^2\tau} \quad(\alpha\tau< 1),
\label{vol}
\end{equation}
where the faster decay is given by the diffusive coefficient $k/m$ and the slow decay of the volatility is provided by the reversion coefficient $\alpha$ (cf. Eq.~(\ref{dexpou})).

Therefore, we conclude that the expOU model appears to be as more realistic. Further investigations on this model are required to confirm this assertion. However, an alternative choice is to sophisticate the OU and Heston models. A second time scale can be generated by including a third SDE for $m$. In a previous work~\cite{pm}, we have studied this possibility for the OU model with some success by taking: $\dot{m}(t)=-\alpha_m(m-\theta)+k_m\xi_3(t)$. A similar analysis can be performed for the Heston model.

\ack

This work has been supported in part by Direcci\'on General de Proyectos de Investigaci\'on under contract No. BFM2003-04574, and by Generalitat de Catalunya under contract No. 2001SGR-00061.

\end{document}